\documentclass[preprint,aps,preprintnumbers,amsmath,amssymb]{revtex4}
\usepackage{graphicx}
\usepackage{dcolumn}
\usepackage{bm}
\usepackage{natbib}

\newcommand{\beq}{\begin{equation}}
\newcommand{\eq}{\end{equation}}
\newcommand{\bea}{\begin{eqnarray}\displaystyle}
\newcommand{\ea}{\end{eqnarray}}
\newcommand{\D}{\nabla}

\begin{document}

\preprint{}

\title{Cubic interactions in the BMN limit of $AdS_3 \times S^3$}
\author{Sera Cremonini}
\email{sera@het.brown.edu}
\author{Aristomenis Donos}
\email{donos@het.brown.edu}
\affiliation{Department of Physics, Brown University \\
Providence, RI 02912, USA}

\date{\today}

\begin{abstract}
We study the pp limit of $AdS_3 \times S^3$ at the interaction level.
We find the interacting Hamiltonian for the bosonic fields of $D=6$ SUGRA in the pp-wave background, and
compare it to the cubic couplings of the full $AdS_3 \times S^3$.
We show how the pp-wave theory vertex arises in the large $J$ limit.
Our analysis also provides some insight into the origin of specific ``prefactors'' which appear in the
pp-wave interaction.
\end{abstract}

\maketitle

\section{Introduction}
The correspondence \cite{Maldacena:1997re,Gubser:1998bc,Witten:1998qj} between supergravity (SUGRA) on Anti-de Sitter (AdS)
space times a compact manifold, and conformal field theory (CFT) living on the boundary of AdS has been
a topic of great interest \cite{Romans:er,Strominger:1996sh,Freedman:1998tz,Maldacena:1998bw,Deger:1998nm,
Balasubramanian:1998sn,Lee:1998bx,Balasubramanian:1998de,deBoer:1998us,Jevicki:1998rr,Corrado:1999pi,Seiberg:1999xz,
Larsen:1999uk,Nastase:1999cb,Bastianelli:1999en,Arutyunov:1999en,Lee:1999pj,Jevicki:1998bm,
Mihailescu:1999cj,Lunin:2000yv,Theisen,Nicolai:2001ac}.
A large number of checks have established this correspondence for a variety of specific cases with the notion of
holography playing a central role.

Recently the AdS/CFT duality was extended to full string theory by Berenstein, Maldacena and Nastase (BMN) who gave
a precise dictionary between the two dual theories \cite{Berenstein:2002jq}.
Concretely, the dual description of IIB string theory on the ten-dimensional pp-wave is the large
R-charge sector of the ${\cal N}=4$ $SU(N)$ gauge theory.
It has been known for some time that in the so-called Penrose limit \cite{penrose}, any spacetime which solves the Einstein field
equations is reduced to a plane-wave background. This assertion has been extended to supergravity backgrounds \cite{Gueven:2000ru}.
In particular, it has been shown that maximally supersymmetric pp-wave backgrounds can be obtained as Penrose limits
of $AdS_p \times S^q$ backgrounds in ten-dimensional IIB SUGRA and eleven-dimensional supergravity
\cite{blau,Kowalski-Glikman:wv,Meessen:2001vx}.
Remarkably, pp-waves provide exact backgrounds for string theory in which the Green-Schwarz worldsheet action becomes
quadratic in the lightcone gauge \cite{Metsaev:2001bj,Metsaev:2002re,Russo:2002rq}.
In the large $J$ limit BMN succeded in reproducing the string spectrum from perturbative Yang-Mills theory.
There is a very clear understanding of the limit from AdS space to the pp-wave background at the classical level.
In particular, the symmetries of $AdS_p \times S^q$ spacetimes contract to the corresponding symmetries of the pp-wave background.
This phenomenon was investigated at the linearized level in \cite{Das:2002cw,Das:2002ij,Arutyunov:2002xd}.

The problem of reconstructing the full interacting string theory
from the large J limit of Yang-Mills represents a definite
challenge. Success was achieved in computing anomalous dimensions
and identifying elements of three-string interactions
\cite{Spradlin:2002ar,Kristjansen:2002bb,Berenstein:2002sa,Gross:2002su,Constable:2002hw,Schwarz:2002bc,
Pankiewicz:2002gs,Vaman:2002ka,Pearson:2002zs,Gomis:2002wi,Kiem:2002pb,Pankiewicz:2002tg,He:2002zu,Kiem:2002xn,deMelloKoch:2003pv}.
Issues related to the question of ``holography'' and of the analog
of the GKP-W formula \cite{Gubser:1998bc,Witten:1998qj} were also
addressed
\cite{Kiritsis:2002kz,Leigh:2002pt,Lunin:2002fw,Asano:2003xp,Mann:2003qp}.
At the level of interactions, a holographic map in the pp-wave
background is still an outstanding problem. In fact, while in the
AdS/CFT correspondence the CFT correlators are recovered from
string theory by using bulk-to-boundary propagators and bulk SUGRA
interaction vertices, in the pp-wave limit the original boundary
is lost, and the notion of the bulk-to-boundary propagator itself
becomes unclear.

In an attempt to shed more light on some of these problems, in the present work we will study the pp-wave limit
of $AdS_3 \times S^3$ at the level of cubic interactions. Our concern is the interacting Hamiltonian for the bosonic fields
of $D=6$ SUGRA in the pp-wave background,
which we will compare with the cubic couplings found in the full $AdS_3 \times S^3$ of \cite{Theisen}.
The comparison will be performed by explicitly taking the large $J$ limit of the cubic form factors constructed in the AdS Hamiltonian.
It will be demonstrated how the pp-wave theory vertex arises in this limit.
This study provides insight into the origin of specific ``prefactors'' appearing in the pp-wave interaction, and hopefully will also help
bring some clarity to the relevance of ``holography'' in the pp limit.

The structure of our paper is the following.
In Section II we give a brief summary of the relevant equations governing $D=6$ SUGRA, as well as discuss the form of the
AdS Hamiltonian. We also list the cubic couplings for the bosonic fields found in the full $AdS_3 \times S^3$
analysis of \cite{Theisen}. In Section III we present our calculations. We work in the light-cone gauge and calculate
the interacting Hamiltonian. We introduce complex chiral primary fields and list the cubic couplings in terms of these.
Finally, in Section IV we compare our vertices to those of \cite{Theisen}, after expressing our pp-wave quantum numbers
in terms of those of the full $AdS_3 \times S^3$, in an appropriate limit. We conclude with a brief summary.

\section{D=6 Supergravity on $AdS_3 \times S^3$}
In order to obtain the cubic couplings of chiral primaries one needs to consider the quadratic corrections to the
covariant equations of motion for $D=6$, ${\cal N}=4b$ supergravity (SUGRA) coupled to $n$ tensor multiplets.
The fields that we will concentrate on are the metric $g_{MN}$, two-forms
$B^i_{MN}, B^r_{MN}$ and scalars. Here $i=1,...,5$ are $SO(5)$ indices and $r=1,...,n$ are $SO(n)$
indices. The scalars are packaged in a $SO(5,n)$ matrix $V_I^J,$ with
$I,J=1,...,5+n$, parametrizing the coset $\frac{SO(5,n)}{SO(5)\times SO(n)}$.
In our notation $M,N$ are $D=6$ coordinates.
The following quantities will be useful in what follows:
\bea
dV\,V^{-1}&=&\begin{pmatrix}
  Q^{ij} & \sqrt{2}P^{is} \\
  \sqrt{2}P^{rj} & Q^{rs}
\end{pmatrix}\\
H^i&=&G^IV^i_I, \ \ H^r=G^IV^r_I, \\
D_MP_N^{ir}&=&\D_M P^{ir}_N - Q^{ij}_M P^{jr}_N-Q^{rs}_M P^{is}_N,
\ea
where $G=dB$. In the bosonic sector, the $D=6$ supergravity equations are the
Einstein equation
\beq
\label{Einstein}
R_{MN}=\frac{1}{2!}H^i_{MPQ}H^{i \ \ PQ}_N + \frac{1}{2!}H^r_{MPQ}H^{r \ \ PQ}_N +2P^{ir}_M P^{ir}_N,
\eq
the scalar equation
\beq
\label{scalareq}
D^MP_M^{ir}=\frac{\sqrt{2}}{3} H^{i}_{MNP} {H^{r}}^{MNP},
\eq
and the Hodge-duality conditions on the 3-form field strengths
\beq
\label{dualityconds}
\ast \, H^i=H^i, \ \ \ast \, H^r=-H^r.
\eq

After computing the quadratic corrections to the above covariant equations of motion, the authors
of \cite{Mihailescu:1999cj} and \cite{Theisen} derived the Lagrangian for the scalar chiral primaries
of the full $AdS_3 \times S^3$ theory.
The cubic interactions of such chiral fields contained derivative terms. However, in a
manner analogous to \cite{Lee:1998bx}, the derivatives were eliminated by performing field
redefinitions and using the on-shell mass condition.
To discuss the general form of the interaction Lagrangian,
consider a complex scalar field $\Phi$ living in $AdS_3 \times S^3$. Since the $AdS_3 \times S^3$
wavefunction is factorizable, one can expand $\Phi$ in terms of spherical harmonics,
\beq
\label{expansion}
\Phi(x,y)=\sum_{I} \Psi_I(x) Y^I(y),
\eq
where $x$ and $y$ denote $AdS_3$ and $S^3$ coordinates respectively, and $I$ refers
to all the $S^3$ quantum numbers.
The interacting Lagrangian is then given by
\beq
\label{L3pp}
L_3= \sum_{I_1,\, I_2,\, I_3} V_{I_1 \, I_2 \, I_3} \int_{AdS_3}
\, \bar{\Psi}_{I_1} \Psi_{I_2}\Psi_{I_3} \int_{S_3} \bar{Y}^{I_1} Y^{I_2} Y^{I_3} + h.c.
\eq
where $V_{I_1 \, I_2 \, I_3}$ contains the energy factors that
resulted from replacing the derivative terms in the way described above.
Letting
$a_{I_1\, I_2\, I_3}=\int_{S^3}\bar{Y}^{I_1} Y^{I_2} Y^{I_3}$, the above integral becomes
\beq
L_3=\sum_{I_1,\, I_2,\, I_3} a_{I_1\, I_2\, I_3} \, V_{I_1 \, I_2 \, I_3} \,  \int_{AdS_3} \,
\bar{\Psi}_{I_1} \Psi_{I_2}\Psi_{I_3} + h.c.
\eq
Thus, the complete Lagrangian for the complex, scalar field $\Psi$ living in $AdS_3$ is
\beq
\label{fullLagrangian}
L= \int_{AdS_3} \sqrt{-g} \, \, \Bigl[ \, \, \sum_I c_1(I) \nabla_{\mu} \bar{\Psi}_I \nabla^{\mu} \Psi_I +
\sum_{I_1, \, I_2, \, I_3} \Bigl( c_2(I_1,I_2,I_3) \bar{\Psi}_{I_1} \Psi_{I_2} \Psi_{I_3} +h.c. \Bigr) \Bigr]
\eq
where the factor $c_1(I)$ represents
the integral over two spherical harmonics, and
$c_2(I_1,I_2,I_3)=a_{I_1\, I_2\, I_3} \, V_{I_1 \, I_2 \, I_3}$.

We can now list the interaction terms found in \cite{Theisen} for
the full $AdS_3 \times S^3$ theory.
In the notation of \cite{Theisen}, the chiral primaries are
denoted by $\sigma$, a singlet with respect to the internal symmetry group $SO(n)$, and
$s^r$, transforming in the fundamental representation of $SO(n)$.
The cubic couplings which involve the chiral primaries are
\beq
\label{vertices}
{\cal L}^{s^r}(\psi)=V^{s^rs^r\psi}_{I_1 I_2 I_3}s^r_{I_1}s^r_{I_2}\psi_{I_3}, \; \; \;
{\cal L}^{\sigma}(\psi)=V^{\sigma \sigma \psi}_{I_1 I_2 I_3}\sigma_{I_1}\sigma_{I_2}\psi_{I_3}, \; \; \;
{\cal L}^{s^r \sigma}(t^r)=V^{s^rt^r\sigma}_{I_1 I_2 I_3}s^r_{I_1}t^r_{I_2}\sigma_{I_3},
\eq
with $\psi\epsilon \{ \sigma,\tau \}$, and where the scalar fields $\tau$ and $t^r$ are descendents of the chiral primaries.
The vertices appearing in (\ref{vertices}) are
\bea
\label{vertices}
V^{s^rs^r\sigma}_{I_1I_2I_3}&=&\frac{-2^4\Sigma(\Sigma+2)(\Sigma-2)\,
\beta_1 \beta_2 \beta_3}{j_3+1} \, a_{I_1I_2I_3} \nonumber \\
V^{s^rs^r\tau}_{I_1I_2I_3}&=&\frac{2^6(\Sigma+2)(\beta_1+1)(\beta_2+1)
\beta_3(\beta_3-1)(\beta_3-2)}{j_3+1} \, a_{I_1I_2I_3} \nonumber \\
V^{\sigma\sigma\sigma}_{I_1I_2I_3}&=&-\frac{2^3\Sigma(\Sigma+2)(\Sigma-2)\, \beta_1\beta_2\beta_3}
{3(j_1+1)(j_2+1)(j_3+1)} \,(j_1^2+j_2^2+j_3^2-2) \, a_{I_1I_2I_3} \nonumber \\
V^{\sigma\sigma\tau}_{I_1I_2I_3}&=&\frac{2^5(\Sigma+2)\,(\beta_1+1)(\beta_2+1)\beta_3(\beta_3-1)(\beta_3-2)}
{(j_1+1)(j_2+1)(j_3+1)}\,(j_1^2+j_2^2+(j_3+2)^2-2)\, a_{I_1I_2I_3} \nonumber \\
V^{s^rt^r\sigma}_{I_1I_2I_3}&=&2^7\frac{(\Sigma+2)\,(\beta_1+1)\beta_2(\beta_2-1)(\beta_3+1)(\beta_2-2)}
{(j_3+1)}\, a_{I_1I_2I_3}
\ea
where
\beq
\label{betas}
\beta_1 = (j_2+j_3-j_1), \ \ \beta_2=(j_1+j_3-j_2), \ \ \beta_3=(j_1+j_2-j_3),
\eq
and $j$ is the $SO(4)$ angular momentum appearing in the Casimir as $j(j+2)$.

As we will show in the next section, in our study of the pp limit of $AdS_3 \times S^3$,
by adopting the light-cone gauge and going to momentum
space we were able to replace the derivatives in the interaction terms by appropriate
energy factors, without resorting to the use of the on-shell mass condition.
Thus, we are now justified in discussing the form of the AdS Hamiltonian for our
pp-wave analysis.

We start by rescaling the complex field $\Psi$ of (\ref{fullLagrangian}) so as to eliminate the constant $c_1$ from the kinetic term,
$\Psi \rightarrow \sqrt{c_1} \, \Psi$ and $\bar{\Psi} \rightarrow \sqrt{c_1} \, \bar{\Psi}$.
We then introduce momenta $\Pi_{\Psi}=\frac{1}{\sqrt{-g}g^{00}}\frac{\partial L}{\partial \dot{\Psi}}$ and
$\bar{\Pi}_{\Psi}=\frac{1}{\sqrt{-g}g^{00}}\frac{\partial L}{\partial \dot{\bar{\Psi}}}$.
Notice that $\Pi_{\Psi}$ is related to the canonical momentum $\Pi_\Psi^c$ by
$\Pi_{\Psi}^c=\sqrt{-g} \, g^{00} \, \Pi_{\Psi}$.
In terms of $\Pi_{\Psi}$ and $\bar{\Pi}_{\Psi}$ the Hamiltonian $H=\Pi^c_{\Psi}\dot{\Psi}+\bar{\Pi}^c_{\Psi}\dot{\bar{\Psi}}-L$
becomes
\beq
H=\int_{AdS_3}\sqrt{-g}\, [\, g^{00}(\Pi_\Psi \, \bar{\Pi}_\Psi - \partial_i \bar{\Psi} \partial^i \Psi)
- \frac{c_2}{{c_1^{3/2}}} \, \bar{\Psi} \Psi \Psi]
\eq
where the $I$ indices have been suppressed for convenience.

Next, we canonically quantize the fields $\Psi$ and $\Pi$ and their hermitian conjugates.
We let
\bea
\Psi&=& \sum_{\omega,\, l,\, m} \frac{1}{\sqrt{2\omega_{lm}}}(A+B^\dag)_{\omega lm}\Psi_{\omega lm}, \; \;
\bar{\Psi}=\sum_{\omega,\, l,\, m} \frac{1}{\sqrt{2\omega_{lm}}}(A^\dag+B)_{\omega lm}\bar{\Psi}_{\omega lm} \\
\Pi_{\Psi} &=& i\sum_{\omega,\, l,\, m} \sqrt{\frac{\omega_{lm}}{2}}(B-A^\dag)_{\omega lm}\Psi_{\omega lm}, \; \;
\bar{\Pi}_{\Psi}= i\sum_{\omega,\, l,\, m} \sqrt{\frac{\omega_{lm}}{2}}(A-B^\dag)_{\omega lm}\bar{\Psi}_{\omega lm}
\ea
where ${\omega, \, l, \, m}$ are the three $AdS_3$ quantum numbers.
The creation and annihilation operators obey commutation relations
\beq
[A,A^\dag]=[B,B^\dag]=1, \; \;  [A,A]=[A,B]=[B,B]=[A^\dag,A^\dag]=[A^\dag,B^\dag]=[B^\dag,B^\dag]=0.
\eq
For simplicity we collectively refer to the $AdS_3$ quantum numbers by $\{n\}$.
The quadratic Hamiltonian then becomes
\beq
H_2 = \sum_{ \{n\} } \omega_{n} (B_n^\dag B_n +A_n^\dag A_n)
\eq
where we have used the orhonormality property
\beq
\int_{AdS_3} \sqrt{-g} g^{00} \Psi_{\{n\}} \Psi_{\{n^\prime\}}=\delta_{n,n^\prime}
\eq
and where we have neglected the vacuum energy.
The cubic Hamiltonian can also be expanded in terms of creation and annihilation operators,
\beq
H_3 \sim  - \sum_{\{n\}, \, \{n^\prime\}, \, \{n^{\prime \prime} \} } \int_{AdS_3}
\frac{\sqrt{-g} g^{00}}{\sqrt{\omega_{n} \omega_{n^\prime} \omega_{n^{\prime \prime} }} }
(A^\dag_n+B_n)(A_{n^\prime}+B^\dag_{n^\prime})(A_{n^{\prime \prime}}+B^\dag_{n^{\prime \prime}})
\bar{\Psi}_{\{n\}} \Psi_{\{n^\prime\}} \Psi_{\{n^{\prime \prime} \}}+h.c. \nonumber
\eq
Finally, inserting all the needed factors, the full Hamiltonian has the form
\bea
H &=& H_2 + H_3 \nonumber \\
&=&  \sum_{ \{n\} } \omega_{n} (B_n^\dag B_n +A_n^\dag A_n) - \frac{1}{2} \,
\sum_{I_1, \, I_2, \, I_3}
\, \frac{a_{I_1\, I_2\, I_3}}{c_1^{3/2}} \, f(\alpha_1,\alpha_2,\alpha_3) \,
\, \sum_{\{n\}, \, \{n^\prime\}, \, \{n^{\prime \prime} \} }
\frac{2^{-3/2}}{\sqrt{\omega_{n} \omega_{n^\prime} \omega_{n^{\prime \prime} }} } \,  \times \nonumber \\
&&
\times \, \biggl( \, \, \int_{AdS_3} \sqrt{-g} g^{00} \bar{\Psi}^{I_1}_{\{n\}} \Psi^{I_2}_{\{n^\prime\}} \Psi^{I_3}_{\{n^{\prime \prime} \}}
(A^\dag_n+B_n)(A_{n^\prime}+B^\dag_{n^\prime})(A_{n^{\prime \prime}}+B^\dag_{n^{\prime \prime}})+h.c.\biggr) \nonumber
\ea

\section{$AdS_3 \times S^3$ in the pp light cone gauge}

In this section we compute the quadratic corrections to the equations of motion for the bosonic fields of $D=6$ SUGRA in the pp-limit.
We adopt the light-cone gauge, and find the interaction Hamiltonian.
Our analysis is similar to that of \cite{Kiem:2002pb} for the pp-limit of $AdS_5 \times S^5$.

\subsection{The background}
We choose to work using light-cone coordinates $(x^+,x^-,x_I)$,
where the transverse directions $x_I$
are denoted by the $SO(4)$ index $I=1,...,4$. The coordinates $x^+$ and $x_-$ are built out of the $AdS_3$ time
coordinate and the corresponding $S^3$ angle, while $x_I$
parametrize the remaining $4$-dimensional $SO(4)$-invariant transverse space.
Since we want to obtain the pp-wave as a solution, we take the background field strength to be
\beq
\label{backgd}
\bar{G}_{+12}=\sqrt{2}\mu \, \epsilon_{+12},\ \ \bar{G}_{+34}=\sqrt{2}\mu
\epsilon_{+34} \ \ \ (\epsilon_{+12}=\epsilon_{+34}=1),
\eq
and we take $V_I^J=\delta_I^J$.
This gives $\bar{H}^5_{+12}=\bar{H}^5_{+34}=\sqrt{2}\mu,$
while the remaining components of $H$ vanish.
Thus, the only nontrivial component of the Einstein equations is
$$R_{++}=4\mu^2,$$
whose solution is the background pp-wave metric
\beq
\label{metric}
ds^2=2\,dx^{+}dx^{-}-\mu^2x^2_I\,dx^{+}dx^{+}+dx^Idx^I.
\eq
Note that the background field strength breaks the
$SO(4)$ symmetry to $SO(2)_{\parallel}\times SO(2)_\perp$, where the parallel directions are along $AdS_3$,
while the perpendicular ones are along the $S^3$.

\subsection{The Fluctuations}
We choose the metric fluctuations to be parametrized as follows,
\beq
g_{\mu\nu}=\begin{pmatrix}
  g_{++} & g_{+-} & g_{+J} \\
  g_{-+} & g_{--} & g_{-J} \\
  g_{I+} & g_{I-} & g_{IJ}
\end{pmatrix},
\eq
where
\beq
\label{metricfluctuations}
g_{++}=-\mu^2 \, x^2_I + h_{++}, \ \ \ g_{+-}=e^{\varphi}, \ \ \ g_{IJ}=e^{\psi}\gamma_{IJ},
\eq
and det$(\gamma_{IJ})=1$. We choose to work in the light-cone
gauge, and use five out of the six available gauge degrees of
freedom to set $g_{--}=g_{-I}=0$. The remaining gauge invariance is used to impose a relation between $g_{+-}$
and det($g_{\mu\nu}$), $$\varphi=\frac{\psi}{2}.$$
Finally, we keep only one of the 10 physical graviton degrees
of freedom, and let
\beq
\label{gamma}
\gamma_{IJ}=\begin{pmatrix}
  e^h I_2 & 0 \\
  0 & e^{-h} I_2
\end{pmatrix}.
\eq

The fluctuations in the field strength are parametrized as
\bea
\label{formsfluctuations}
G^I &=& \bar{G}^I+g^I, \ \ g^I_{MNP}=3\partial_{[M}b^I_{NP]}, \\
V_I^i &=& \delta_I^i+\Phi^{ir}\delta_I^r+\frac{1}{2}\Phi^{ir}\Phi^{jr}\delta_I^j, \\
V_I^r &=&
\delta_I^r+\Phi^{ir}\delta_I^i+\frac{1}{2}\Phi^{ir}\Phi^{is}\delta_I^s.
\ea
This gives
\bea
\label{Hfluctuations}
H^{i} &=& g^{i}+ \delta^{i5} \bar{G}^{5}+ \phi^{ir}g^{r}+\frac{1}{2}\bar{G}^{5}\phi^{5r}\phi^{ir}, \\
H^{r} &=& g^{r}+ \tilde{G}^{5}\phi^{5r}+\phi^{ir}g^{i}.
\ea
We impose the light-cone gauge condition $b^I_{-N}=0$ on the
2-form gauge fields. Also, we turn on fluctuations that are
scalars under $SO(2)_{\parallel} \times SO(2)_{\perp}$
transformations; as a result, all mixed components of the 2-form vanish, $b_{ij'}=0,$
where $i=1,2$ are $SO(2)_{\parallel}$ vector indices and $i'=3,4$ are $SO(2)_{\perp}$ vector
indices. 

\subsection{Solving for the auxiliary fields}
To express the auxiliary fields $\psi,g_{+I},h_{++}$ in terms of the physical fields, we
will consider appropriate components of the Einstein equation.
The $(--)$ component of (\ref{Einstein}) gives
\beq
\label{R--}
R_{--} = 2 \partial^2_{-} \psi +(\partial_{-} h)^2=
(\partial_{-} b^I_{12})^2+(\partial_{-} b^I_{34})^2 + (\partial_{-} \phi^{ir})^2 + {\cal O} \bigl(3\bigr)
\eq
yielding, up to quadratic level,
\beq
\label{psi}
\psi=\frac{1}{2} \frac{1}{\partial_{-}^2}\bigl[-(\partial_{-}h)^2 +
(\partial_{-}b^I_{12})^2+(\partial_{-}b^I_{34})^2+(\partial_{-}\phi^{ir})^2\bigr].
\eq
The inverse derivative operator is understood to mean
\beq
\frac{1}{\partial_-}g(x^-)=\int^{x^-} dx^{- \; \prime} g(x^{- \; \prime}).
\eq
The $(+-)$ component of (\ref{Einstein}) gives
\beq
\label{R+-}
R_{+-}=-\frac{1}{2} \bigl( \partial^2_{-} h_{++}+ \partial_{-} \partial^J g_{+J}+...\bigr)=
\sqrt{2} \mu \partial_{-}(b^5_{12}+b^5_{34})+{ \cal O} \bigl( 2 \bigr).
\eq
Finally, the $(-J)$ component of the Einstein equation will allow us to express
$g_{+J}$ in terms of $h$.
The right-hand side of (\ref{Einstein})
is zero to linear order in the fields, while the left-hand side is given by
\beq
\label{R-J}
R_{-J}=-\frac{1}{2} \partial^2_- \bigl( g_{+J} + \Delta_{JK} \frac{\partial_K}{\partial_-}h\bigr), \ \ \ \text{with}
\ \ \ \Delta_{JK}=\begin{pmatrix}
  I_2 &  \\
   & -I_2
\end{pmatrix}.
\eq
Solving (\ref{R-J}) for $g_{+J}$, we get
\beq
\label{g+J}
g_{+J}=-\Delta_{JK} \frac{\partial_k}{\partial_-}h +{\cal O}\bigl( 2 \bigr).
\eq
Plugging $g_{+J}$ back into $R_{+-}$, one obtains the following equation for $h_{++}$, to linear order
\beq
\label{h++}
h_{++}^{(1)}=-2^{3/2}\mu \frac{1}{\partial_-}(b^5_{12}+b^5_{34})+
\bigl(\frac{\partial^2_j}{\partial^2_-}-\frac{\partial^2_{j \, '}}{\partial^2_-}\bigr)h +{ \cal O} \bigl( 2 \bigr),
\eq
where $j=1,2$ and $j^{\, \prime}=3,4$.
Then to linear order $g^{--}$ is given by
\bea
g^{--}&=&\mu^2x^2_I-h_{++} \\
&=&\mu^2x^2_I+2^{3/2}\mu \frac{1}{\partial_-}(b^5_{12}+b^5_{34})-
\bigl(\frac{\partial^2_j}{\partial^2_-}-\frac{\partial^2_{j \, '}}{\partial^2_-}\bigr)h.
\ea

\subsection{Solving the duality conditions}
Next, we will use the self-duality and anti-self-duality conditions to obtain the equations of motion for
the physical degrees of freedom of the two-forms.
From the $(-12)$ component of the duality conditions (\ref{dualityconds}) we obtain
\bea
\label{duality-12}
\partial_{-}b^{r}_{12} &=& \left(1+2h\right)\partial_{-}b^{r}_{34}+2\partial_{-}b^{i}_{34}\phi^{ir}, \\
\partial_{-}b^{i}_{12} &=&
-\left(1+2h\right)\partial_{-}b^{i}_{34}-2\partial_{-}b^{r}_{34}\phi^{ir},
\ea
which gives us the following expressions in terms of the
physical fields $a^{i}$ and $a^{r}$,
\bea
b^{r}_{12} &=& a^{r}+{1 \over \partial_{-}}\left(h \partial_{-}a^{r}\right)-
{1 \over \partial_{-}}\left(\phi^{ir} \partial_{-}a^{i}\right), \\
b^{r}_{34} &=& a^{r}-{1 \over \partial_{-}}\left(h
\partial_{-}a^{r}\right)+ {1 \over \partial_{-}}\left(\phi^{ir}\partial_{-}a^{i}\right), \\
b^{i}_{12} &=& a^{i}+{1 \over \partial_{-}}\left(h \partial_{-}a^{i}\right)-
{1 \over \partial_{-}}\left(\phi^{ir} \partial_{-}a^{r}\right), \\
b^{i}_{34} &=& -a^{i}+{1 \over \partial_{-}}\left(h\partial_{-}a^{i}\right)- {1 \over \partial_{-}}\left(\phi^{ir}
\partial_{-}a^{r}\right).
\ea
The $(-+M)$ components of the duality conditions will help us determine the $b_{+M}$ components;
for example, the $(-+1)$ component yields
\bea
\partial_{-}b_{+1}^{r}&=&g^{2-}\partial_{-}b_{34}^{r}+2
\partial_{2}b_{34}^{i} \phi^{ir} + \left(1+h\right)\partial_{2}b_{34}^{r}, \nonumber \\
-\partial_{-}b_{+1}^{i}&=&g^{2-}\partial_{-}b_{34}^{i}+2\partial_{2}b_{34}^{r} \phi^{ir} +
\left(1+h\right)\partial_{2}b_{34}^{i}. \nonumber
\ea
Using the above we find
\bea
b^{i}_{+I}&=&-\epsilon_{IJ} {1 \over \partial_{-}}
[-\left(1+h\right)\partial_{J}a^{i}+{\partial_{J} \over
\partial_{-}}\left(h\partial_{-}a^{i}\right)-{\partial_{J} \over
\partial_{-}}\left(\phi^{ir}\partial_{-}a^{r}\right) +2 \partial_{J}a^{r}
\phi^{ir}-g^{-J}\partial_{-}a^{i}], \nonumber \\
b^{i}_{+I^{\prime}}&=&-\epsilon_{I^{\prime}J^{\prime}} {1 \over\partial_{-}}
[\left(1-h\right)\partial_{J^{\prime}}a^{i}+{\partial_{J^{\prime}}
\over\partial_{-}}\left(h\partial_{-}a^{i}\right)-{\partial_{J^{\prime}} \over
\partial_{-}}\left(\phi^{ir}\partial_{-}a^{r}\right) +2 \partial_{J^{\prime}}a^{r}
\phi^{ir}+g^{-J^{\prime}}\partial_{-}a^{i}], \nonumber \\
b^{r}_{+I}&=&\epsilon_{IJ} {1 \over \partial_{-}}
[\left(1+h\right)\partial_{J}a^{r}-{\partial_{J} \over
\partial_{-}}\left(h\partial_{-}a^{r}\right)+{\partial_{J} \over
\partial_{-}}\left(\phi^{ir}\partial_{-}a^{i}\right) -2 \partial_{J}a^{i}
\phi^{ir}+g^{-J}\partial_{-}a^{r}], \nonumber \\
b^{r}_{+I^{\prime}}&=&\epsilon_{I^{\prime}J^{\prime}} {1 \over
\partial_{-}}[\left(1-h\right)\partial_{J^{\prime}}a^{r}+{\partial_{J^{\prime}}
\over\partial_{-}}\left(h\partial_{-}a^{r}\right)-{\partial_{J^{\prime}} \over
\partial_{-}}\left(\phi^{ir}\partial_{-}a^{i}\right) +2 \partial_{J^{\prime}}a^{i}
\phi^{ir}+g^{-J^{\prime}}\partial_{-}a^{r}]. \nonumber
\ea
The scalar equation then becomes, at second order in fluctuations,
\beq
\label{scalareom}
{1 \over 2}{1 \over \sqrt{g}} \partial_{M}\left( \sqrt{g} g^{MN}
\partial_{N} \phi^{ir} \right) =2\sqrt{2}\mu\delta^{i5}\partial_{-}a^{r}-
2\partial^{I}a^{i}\partial_{I}a^{r}+\partial_{-}a^{i}{\partial^{2}_{I}
\over \partial_{-}}a^{r}+\partial_{-}a^{r}{\partial^{2}_{I} \over
\partial_{-}}a^{i}-[I\longleftrightarrow I^{\prime}].
\eq
The equations of motion for the physical fields $a^{i}$ and
$a^{r}$ are obtained by using
\beq
\partial_{-}[ \left(\ast H^{i}\right)_{+12}-\left(\ast
H^{i}\right)_{+34}]=\partial_{-}
[H^{i}_{+12}-H^{i}_{+34}]
\eq
and
\beq
\partial_{-}[ \left(\ast H^{r}\right)_{+12}+\left(\ast
H^{r}\right)_{+34}]=-\partial_{-}
[H^{r}_{+12}+H^{r}_{+34}].
\eq
The conditions above lead to the following sets of equations, up to quadratic level:
\bea
\label{aeom}
&\nabla^2 a^{i}-2\sqrt{2}\mu\partial_{-}h\delta^{i5}={\partial^{2}_{I}
\over\partial_{-}}\left(h\partial_{-}a^{i}\right)-{\partial^{2}_{I} \over
\partial_{-}}\left(\phi^{ir}\partial_{-}a^{r}\right)+2\partial^{I}\left(\partial_{I}a^{r}\phi^{ir}\right)
-\partial_{I}\left(\partial_{-}a^{i} {\partial^{I} \over \partial_{-}}h\right)\nonumber \\
&-\partial^{I}\left(h\partial_{I}a^{i}\right)+\partial_{-}\left( {\partial^{2}_{I} \over
\partial^{2}_{-}}h\partial_{-}a^{i}\right)
-\partial_{-}\left(\partial_{I}a^{i}{\partial_{I} \over
\partial^{2}_{-}}h\right)
-\partial_{-} \left(\phi^{ir} {\partial^{2}_{I} \over
\partial_{-}}a^{r}\right)
+ \partial_{-} \left(h {\partial^{2}_{I} \over
\partial_{-}}a^{i}\right), \nonumber \\
\nonumber \\
&\nabla^2 a^{r}+2\sqrt{2}\mu \partial_{-} \phi^{5r}= {\partial^{2}_{I} \over
\partial_{-}}\left(h\partial_{-}a^{r}\right)-{\partial^{2}_{I} \over
\partial_{-}}\left(\phi^{ir}\partial_{-}a^{i}\right)+2\partial^{I}\left(\partial_{I}a^{i}\phi^{ir}\right)
-\partial_{I}\left(\partial_{-}a^{r} {\partial^{I} \over\partial_{-}}h\right)\nonumber \\
&-\partial^{I}\left(h\partial_{I}a^{r}\right)+
\partial_{-}\left( {\partial^{2}_{I} \over \partial^{2}_{-}}h\partial_{-}a^{r}\right)
-\partial_{-}\left(\partial_{I}a^{r}{\partial_{I} \over\partial^{2}_{-}}h\right)
-\partial_{-} \left(\phi^{ir} {\partial^{2}_{I} \over
\partial_{-}}a^{i}\right)
+\partial_{-} \left(h {\partial^{2}_{I} \over
\partial_{-}}a^{r}\right),
\ea
where
$$
\nabla^2=2\partial_+\partial_- + \partial_I^2 +\mu^2x_I^2\partial^2_-.
$$

\subsection{The gravity sector}
Since all the terms which contain $a^i, a^r, \phi^{ir}$ have been found,
up to quadratic order at the level of equations of motion, for the
gravity sector one can set $a^i=a^r=\phi^{ir}=0$ and consider pure gravity.
The terms containing $h$ alone can in fact be found by expanding the action
\beq
S_G=-\frac{1}{2\kappa^2} \int d^6x \sqrt{-g}R
=-\frac{1}{2\kappa^2} \int d^6x(2g^{+-}R_{+-}+2g^{I-}R_{I-}+g^{--}R_{--}+g^{IJ}R_{IJ})
\eq
to cubic order. We will follow closely the analysis of \cite{Goroff:hc} adapting their work to our pp-wave background.
Recall that $\psi$ and $g_{+I}$ were found by solving, respectively, $R_{--}=0$ and $R_{I-}=0$,
with all sources set to zero. Thus, these terms don't contribute to the Ricci scalar $R$.
$R_{+-}$, however, contains surface terms that vanish in the action, and its contribution must therefore be considered.
Using $R_{--}=0$ and $R_{I-}=0$ the gravity action reduces to
\beq
\label{GRaction}
S_G=-\frac{1}{2\kappa^2} \int d^6x(2g^{+-}R_{+-}+g^{IJ}R_{IJ}).
\eq
After expanding the action to cubic order one can see that the only term containing $\mu$ is quadratic in $h$.
Thus, the the $\mu$-dependence does not affect the interactions, allowing one to use the results of \cite{Goroff:hc}
to write down the cubic terms.
We find the following terms in the Lagrangian,
\bea
\label{hterms}
L_{hh}&=&h \nabla^{2}h,  \nonumber \\
L_{hhh}&=&2\left(h \partial_{-}h{\partial^{2}_{I} \over \partial_{-}}h- {1
\over 2} \partial^{I}h\partial_{I}hh+ {1\over 2}
\partial_{-}h\partial_{-}h{\partial^{2}_{I} \over
\partial^{2}_{-}}h- \partial_{-}h\partial^{I}h{\partial_{I} \over
\partial_{-}}h\right).
\ea

\subsection{The SUGRA action}

Adding together (\ref{scalareom}),(\ref{aeom}) and (\ref{hterms}), after properly
rescaling some of the terms, we find
\bea
\label{action}
& S= \int d^6x \sqrt{-g}[R-{1 \over \sqrt{g}}\left(
\ast H^{i}\wedge H^{i}+\ast H^{r}\wedge H^{r} \right)+\phi^{ir}
\nabla^{2} \phi^{ir} ] \nonumber \\
& =\int_{R^{6}} dx^{+}dx^{-}d^{4}x  \Bigl[{1 \over
4}\phi^{ir}\partial_{M}\left( g^{MN}
\partial_{N} \phi^{ir} \right)+{1 \over 2} a^{i} \nabla^2 a^{i}+{1 \over 2} a^{r}
\nabla^2 a^{r} -2\sqrt{2}\mu a^{5}\partial_{-}h \nonumber \\
&+2\sqrt{2}\mu a^{r}\partial_{-}\phi^{5r}
+\bigl(h\partial_{-}a^{i}{\partial^{2}_{I} \over \partial_{-}}a^{i}- {1
\over 2} \partial^{I}a^{i}\partial_{I}a^{i}h
+{1\over 2}\partial_{-}a^{i}\partial_{-}a^{i}{\partial^{2}_{I} \over
\partial^{2}_{-}}h -\partial_{-}a^{i}\partial^{I}a^{i}{\partial_{I} \over
\partial_{-}}h \nonumber \\
&+2 \partial_{I}a^{i}\partial^{I}a^{r}\phi^{ir} -\phi^{ir}
\partial_{-}a^{i} {\partial^{2}_{I}\over \partial_{-}} a^{r} -\phi^{ir}
\partial_{-}a^{r} {\partial^{2}_{I}\over \partial_{-}} a^{i}
+h\partial_{-}a^{r}{\partial^{2}_{I} \over \partial_{-}}a^{r}- {1
\over 2} \partial^{I}a^{r}\partial_{I}a^{r}h+ \nonumber \\
&{1\over 2}\partial_{-}a^{r}\partial_{-}a^{r}{\partial^{2}_{I} \over
\partial^{2}_{-}}h- \partial_{-}a^{r}\partial^{I}a^{r}{\partial_{I} \over
\partial_{-}}h-[I\longleftrightarrow
I^{\prime}]\bigr)+{1 \over 4}L_{hh}+{1 \over 4}L_{hhh} \Bigr]\nonumber \\
&= \int_{R^{6}} dx^{+}dx^{-}d^{4}x
\Bigl[{1 \over4}\phi^{ir}\nabla^2 \phi^{ir}
+{1 \over 2} a^{i} \nabla^2 a^{i}+{1 \over 2} a^{r}
\nabla^2 a^{r} \nonumber \\
&-2\sqrt{2}\mu a^{5}\partial_{-}h+2\sqrt{2}\mu a^{r}\partial_{-}\phi^{5r}+\bigl(h
\partial_{-}a^{i}{\partial^{2}_{I} \over \partial_{-}}a^{i}- {1
\over 2} \partial^{I}a^{i}\partial_{I}a^{i}h+ {1\over 2}
\partial_{-}a^{i}\partial_{-}a^{i}{\partial^{2}_{I} \over
\partial^{2}_{-}}h- \partial_{-}a^{i}\partial^{I}a^{i}{\partial_{I} \over
\partial_{-}}h \nonumber \\
& +2 \partial_{I}a^{i}\partial^{I}a^{r}\phi^{ir} -\phi^{ir}
\partial_{-}a^{i} {\partial^{2}_{I}\over \partial_{-}} a^{r} -\phi^{ir}
\partial_{-}a^{r} {\partial^{2}_{I}\over \partial_{-}} a^{i}
+h\partial_{-}a^{r}{\partial^{2}_{I} \over \partial_{-}}a^{r}\nonumber \\
&-{1\over 2} \partial^{I}a^{r}\partial_{I}a^{r}h+ {1\over 2}
\partial_{-}a^{r}\partial_{-}a^{r}{\partial^{2}_{I} \over
\partial^{2}_{-}}h- \partial_{-}a^{r}\partial^{I}a^{r}{\partial_{I} \over
\partial_{-}}h+{1 \over 4}
\partial^{I}\phi^{ir}\partial_{I}\phi^{ir}h-
{1\over 2}\partial_{-}\phi^{ir}\partial_{I}\phi^{ir} {\partial^{I}
\over\partial_{-}}h \nonumber \\
&+{1\over 4}\partial_{-}\phi^{ir}\partial_{-}\phi^{ir} {\partial^{2}_{I} \over \partial^{2}_{-}}h
 -[I\longleftrightarrow I^{\prime}]\bigr)+{1 \over
4}L_{hh}+{1 \over 4}L_{hhh} \Bigr].
\ea
We introduce complex chiral fields by forming the following linear field redefinitions
\bea
s^{i}&=&{\delta^{i5} \over \sqrt{2}}h+ i a^{i}, \nonumber \\
\sigma^{r}&=&{1 \over \sqrt{2}} \phi^{5r}+ i a^{r}.
\ea
The quadratic Lagrangian then becomes
\bea
\label{L2}
&L_2=\int dx^+ dx^- d^4x_I
\Bigl[\frac{1}{4}(s^i \nabla^2 \bar{s}^i + \bar{s}^5 \nabla^2 s^5 + \bar{\sigma}^r \nabla^2 \sigma^r +h.c.)
+ \frac{1}{4}\phi^{ir} \nabla^2 \phi^{ir} \nonumber \\
&+i\mu(s^5 \partial_-\bar{s}^5 -\bar{s}^5\partial_-s^5 -\sigma^r \partial_-\bar{\sigma}^r
+\bar{\sigma}^r\partial_-\sigma^r)\Bigr],
\ea
where $i \neq 5$.

Before rewriting the full action in terms of the fields introduced above, we
examine the interaction Hamiltonian that comes from (\ref{action}) in the momentum basis.
As it may be seen the cubic Hamiltonian $H_3=-L_3$ is given by
\bea
\label{H3}
&H_{3}= \int_{R^{6}} dx^{-}d^{4}x \,
\bigl\{-h\partial_{-}a^{i}{\partial^{2}_{I} \over \partial_{-}}a^{i}+ {1
\over 2} \partial^{I}a^{i}\partial_{I}a^{i}h- {1\over 2}
\partial_{-}a^{i}\partial_{-}a^{i}{\partial^{2}_{I} \over
\partial^{2}_{-}}h+ \partial_{-}a^{i}\partial^{I}a^{i}{\partial_{I} \over
\partial_{-}}h \nonumber \\
& -2 \partial_{I}a^{i}\partial^{I}a^{r}\phi^{ir} +\phi^{ir}
\partial_{-}a^{i} {\partial^{2}_{I}\over \partial_{-}} a^{r} +\phi^{ir}
\partial_{-}a^{r} {\partial^{2}_{I}\over \partial_{-}} a^{i} \nonumber \\
& -h\partial_{-}a^{r}{\partial^{2}_{I} \over \partial_{-}}a^{r}+ {1
\over 2} \partial^{I}a^{r}\partial_{I}a^{r}h- {1\over 2}
\partial_{-}a^{r}\partial_{-}a^{r}{\partial^{2}_{I} \over
\partial^{2}_{-}}h+ \partial_{-}a^{r}\partial^{I}a^{r}{\partial_{I} \over
\partial_{-}}h \nonumber \\
& -{1 \over 4} \partial^{I}\phi^{ir}\partial_{I}\phi^{ir}h+
{1\over 2}\partial_{-}\phi^{ir}\partial_{I}\phi^{ir} {\partial^{I}
\over
\partial_{-}}h - {1\over 4}
\partial_{-}\phi^{ir}\partial_{-}\phi^{ir} {\partial^{2}_{I} \over \partial^{2}_{-}}h \nonumber \\
& + {1 \over 2}\left(-h \partial_{-}h{\partial^{2}_{I} \over \partial_{-}}h+ {1
\over 2} \partial^{I}h\partial_{I}hh- {1\over 2}
\partial_{-}h\partial_{-}h{\partial^{2}_{I} \over
\partial^{2}_{-}}h+ \partial_{-}h\partial^{I}h{\partial_{I} \over
\partial_{-}}h \right) -[I\longleftrightarrow I^{\prime}] \, \bigr\} \nonumber \\
&= \int_{R^{6}} dx^{-}d^{4}x \, \bigl\{ {1\over 2}[-h\partial_{-}a^{i}{\partial^{2}_{I} \over
\partial_{-}}a^{i}-{\partial_{I} \over\partial_{-}}\left( h\partial_{-}a^{i}\right) \partial_{I}a^{i} +
2 \partial^{I}a^{i}\partial_{I}a^{i}h] \nonumber \\
&+{1\over 2}[-\partial_{-}a^{i}\partial_{-}a^{i}{\partial^{2}_{I} \over
\partial^{2}_{-}}h- \partial^{I}a^{i}\partial_{I}a^{i}h+ 2 \partial_{-}a^{i}\partial^{I}a^{i}{\partial_{I} \over
\partial_{-}}h] \nonumber \\
& {1\over 4}[-h\partial_{-}h{\partial^{2}_{I} \over
\partial_{-}}h-{\partial_{I} \over
\partial_{-}}\left( h\partial_{-}h\right) \partial_{I}h + 2 \partial^{I}h\partial_{I}hh] \nonumber \\
&+{1\over 4}[ -\partial_{-}h\partial_{-}h{\partial^{2}_{I}
\over\partial^{2}_{-}}h- \partial^{I}h\partial_{I}hh+ 2 \partial_{-}h\partial^{I}h{\partial_{I} \over
\partial_{-}}h] \nonumber \\
& {1\over 2} [-h\partial_{-}a^{r}{\partial^{2}_{I} \over \partial_{-}}a^{r}-{\partial_{I} \over
\partial_{-}}\left( h\partial_{-}a^{r}\right) \partial_{I}a^{r} +  2 \partial^{I}a^{r}\partial_{I}a^{r}h]\nonumber \\
&+{1\over 2}[ - \partial^{I}a^{r}\partial_{I}a^{r}h+ 2
\partial_{-}a^{r}\partial^{I}a^{r}{\partial_{I} \over
\partial_{-}}h-\partial_{-}a^{r}\partial_{-}a^{r}{\partial^{2}_{I} \over
\partial^{2}_{-}}h] \nonumber \\
& {1 \over 4}[ -\partial^{I}\phi^{ir}\partial_{I}\phi^{ir}h+
2\partial_{-}\phi^{ir}\partial_{I}\phi^{ir} {\partial^{I}
\over\partial_{-}}h -
\partial_{-}\phi^{ir}\partial_{-}\phi^{ir} {\partial^{2}_{I} \over \partial^{2}_{-}}h] \nonumber \\
& +[-2\partial^{I}a^{i}\partial_{I}a^{r}\phi^{ir}+
\partial_{-}a^{i} {\partial_{I}^{2} \over
\partial_{-}}a^{r}\phi^{ir} +\partial_{-}a^{r} {\partial^{2}_{I} \over \partial^{2}_{-}}a^{i}\phi^{ir}]-
[I\longleftrightarrow I^{\prime}] \, \bigr\} \nonumber \\
& = H_3^{(1)}+H_3^{(2)}+H_3^{(3)}+H_3^{(4)}+H_3^{(5)}+H_3^{(6)}
\ea
where the terms have been arranged in this way for later convenience.
Before continuing, it's important to make a comment.
The Hamiltonian written above is the light-cone Hamiltonian $H_{lc}=i\partial_+=i(\partial_t+\partial_\theta)$, where $t$ is
the global time coordinate of $AdS_3$, and $\theta$ is the corresponding $S^3$ coordinate.
This differs from the ``global'' $AdS$ Hamiltonian given by $H=i\partial_t$, which was discussed in Section II.
However, in the infinite-momentum frame the two coincide, $i\partial_+ \rightarrow i\partial_t$,
making the discussion in Section II relevant.

We can rewrite the Hamiltonian in momentum space, after performing the replacement
\beq
\label{momentumbasis}
\frac{i}{\partial_-^r}\rightarrow \frac{1}{p^+_r} = \frac{1}{\alpha_r},
\eq
where $r=1,2,3$ denotes particle number.
Thus, by going to the regular Fourier-transformed representations of the fields, we find,
for two functions $f$ and $g$, the following expressions:
\bea
& \int_{R^{6}} dx^{-}d^{4}x
[-f\partial_{-}g{\partial^{2}_{I} \over
\partial_{-}}f-{\partial_{I} \over
\partial_{-}}\left( f\partial_{-}f\right) \partial_{I}g + 2
f\partial^{I}g\partial_{I}f] \nonumber \\
&= \int_{R^{6}}\prod_{r}{d\alpha_{r}d^{4}p_{r} \over
(\sqrt{2\pi})^{5}}\delta\left(\Delta\alpha_{r}\right)
\delta\left(\Delta p_{r}\right)[{-1 \over \alpha_{2}\alpha_{3}}P^{2}
]f_{1}f_{2}g_{3}, \\
\nonumber \\
& \int_{R^{6}} dx^{-}d^{4}x [
-\partial_{-}f\partial_{-}g{\partial^{2}_{I} \over
\partial^{2}_{-}}f- \partial^{I}f\partial_{I}gf+ 2 \partial_{-}f\partial^{I}g{\partial_{I} \over
\partial_{-}}f] \nonumber \\
&= \int_{R^{6}}\prod_{r}{d\alpha_{r}d^{4}p_{r} \over (\sqrt{2\pi})^{5}}
\delta\left(\Delta \alpha_{r}\right)
\delta\left(\Delta p_{r}\right)[-{1 \over 2} \left({1 \over
\alpha_{1}^{2}}+{1 \over \alpha_{2}^{2}}\right)P^{2} ]f_{1}f_{2}g_{3}, \\
\nonumber \\
& \int_{R^{6}} dx^{-}d^{4}x
[-g\partial_{-}f{\partial^{2}_{I} \over
\partial_{-}}f-{\partial_{I} \over
\partial_{-}}\left( g\partial_{-}f\right) \partial_{I}f + 2
\partial^{I}f\partial_{I}fg] \nonumber \\
&= \int_{R^{6}}\prod_{r}{d\alpha_{r}d^{4}p_{r} \over
(\sqrt{2\pi})^{5}}\delta\left(\Delta \alpha_{r}\right)
\delta\left(\Delta p_{r}\right)[{-1 \over \alpha_{1}\alpha_{3}}P^{2}
]f_{1}g_{2}f_{3},
\ea
where $\delta{(\Delta \alpha_r)} \equiv \delta{(\alpha_1+\alpha_2-\alpha_3)}$,
$\delta{(\Delta p_r)} \equiv \delta{(p_1+p_2-p_3)}$,
$P^{2}=\left(\alpha_{1}p_{2}-\alpha_{2}p_{1}\right)^{2}$ and $p_r$ denotes the transverse momentum of the
$r$-th particle.
After applying the general rules given above to the terms in (\ref{H3}), we find, for $i=5$,
\bea
\label{first}
&
H_3^{(1)}=\int_{R^{6}} dx^{-}d^{4}x \bigl\{{1\over 2} [-h\partial_{-}a^{i}{\partial^{2}_{I}
\over\partial_{-}}a^{i}-{\partial_{I} \over\partial_{-}}\left( h\partial_{-}a^{i}\right)
\partial_{I}a^{i} + 2 \partial^{I}a^{i}\partial_{I}a^{i}h ] \nonumber \\
&
+{1\over 2} [-\partial_{-}a^{i}\partial_{-}a^{i}{\partial^{2}_{I} \over
\partial^{2}_{-}}h- \partial^{I}a^{i}\partial_{I}a^{i}h+ 2 \partial_{-}a^{i}
\partial^{I}a^{i}{\partial_{I} \over \partial_{-}}h ] \nonumber \\
&
+{1\over 4} [-h\partial_{-}h{\partial^{2}_{I} \over \partial_{-}}h-{\partial_{I} \over
\partial_{-}}\left( h\partial_{-}h\right) \partial_{I}h + 2 \partial^{I}h\partial_{I}hh] \nonumber \\
&
+{1\over 4}[ -\partial_{-}h\partial_{-}h{\partial^{2}_{I}\over \partial^{2}_{-}}h
-\partial^{I}h\partial_{I}hh+2\partial_{-}h\partial^{I}h{\partial_{I} \over\partial_{-}}h] \bigr\} \nonumber \\
&
=\int_{R^{6}} dx^{-}d^{4}x \, \bigl\{ {1\over 2}[-s^{5}\partial_{-}\bar{s}^{5}{\partial^{2}_{I}
\over\partial_{-}}s^{5}-{\partial_{I} \over
\partial_{-}}\left( s^{5}\partial_{-}\bar{s}^{5}\right) \partial_{I}s^{5} + 2
\partial^{I}s^{5}\partial_{I}\bar{s}^{5}s^{5}] \nonumber \\
&
+{1\over 2}[-\partial_{-}s^{5}\partial_{-}\bar{s}^{5}{\partial^{2}_{I} \over
\partial^{2}_{-}}s^{5}- \partial^{I}s^{5}\partial_{I}\bar{s}^5 s^{5}+ 2 \partial_{-}s^{5}
\partial^{I}\bar{s}^{5}{\partial_{I} \over\partial_{-}}s^{5}] + c.c. \bigr\} \nonumber \\
&
=\int_{R^{6}}\prod_{r}{d\alpha_{r}d^{4}p_{r} \over\sqrt{2\pi}^{5}}\delta\left(\Delta\alpha_{r}\right)
\delta\left(\Delta p_{r}\right)[-{1 \over8}{\alpha^{2}_{1}+\alpha^{2}_{2}+\alpha^{2}_{3}
\over \alpha^{2}_{1}\alpha^{2}_{2}}\, P^{2}]
(s^{5}_{1}s^{5}_{2}\bar{s}^{5}_{3}+c.c.).
\ea
When  $i \neq 5$, instead, we have
\bea
&H_3^{(2)}=\int_{R^{6}} dx^{-}d^{4}x
 \bigl\{ {1\over 2}[-h\partial_{-}a^{i}{\partial^{2}_{I} \over
\partial_{-}}a^{i}-{\partial_{I} \over
\partial_{-}}\left( h\partial_{-}a^{i}\right) \partial_{I}a^{i} + 2 \partial^{I}a^{i}
\partial_{I}a^{i}h] \nonumber \\
&+{1\over 2}[
-\partial_{-}a^{i}\partial_{-}a^{i}{\partial^{2}_{I} \over
\partial^{2}_{-}}h- \partial^{I}a^{i}\partial_{I}a^{i}h+ 2 \partial_{-}a^{i}
\partial^{I}a^{i}{\partial_{I} \over
\partial_{-}}h] \bigr\} \nonumber \\
&=\int_{R^{6}} dx^{-}d^{4}x \bigl\{ {1\over 2}[-s^{5}\partial_{-}\bar{s}^{i}{\partial^{2}_{I}
\over\partial_{-}}s^{i}-{\partial_{I} \over
\partial_{-}}\left( s^{5}\partial_{-}\bar{s}^{i}\right) \partial_{I}s^{i} + 2
\partial^{I}s^{i}\partial_{I}\bar{s}^{i}s^{5}] \nonumber \\
&+{1\over 2}[
-\partial_{-}s^{i}\partial_{-}\bar{s}^{i}{\partial^{2}_{I} \over
\partial^{2}_{-}}s^{5}- \partial^{I}s^{i}\partial_{I}\bar{s}^{i}s^{5}+ 2
\partial_{-}s^{i}\partial^{I}\bar{s}^{i}{\partial_{I} \over
\partial_{-}}s^{5}] + c.c.\bigr\} \nonumber \\
&=\int_{R^{6}}\prod_{r}{d\alpha_{r}d^{4}p_{r} \over
\sqrt{2\pi}^{5}}\delta\left(\Delta \alpha_{r}\right)
\delta\left(\Delta p_{r}\right)\bigl[{1 \over
4}{\alpha^{2}_{2}-\alpha_{1}\alpha_{3} \over \alpha_{1}\alpha_{3}\alpha^{2}_{2}}\, P^{2}\bigr](s^{i}_{1}s^{5}_{2}\bar{s}^{i}_{3}+c.c.).
\ea
The interactions containing $h, a^r$ and $\phi^{5r}$ are:
\bea
& H_3^{(3)}=\int_{R^{6}} dx^{-}d^{4}x {1\over 2}
[-h\partial_{-}a^{r}{\partial^{2}_{I} \over
\partial_{-}}a^{r}-{\partial_{I} \over
\partial_{-}}\left( h\partial_{-}a^{r}\right) \partial_{I}a^{r} +
2\partial^{I}a^{r}\partial_{I}a^{r}h] \nonumber \\
&=\int_{R^{6}}\prod_{r}{d\alpha_{r}d^{4}p_{r} \over
\sqrt{2\pi}^{5}}\delta\left(\Delta \alpha_{r}\right)
\delta\left(\Delta p_{r}\right)\bigl[\frac{-1}{2\alpha_1 \, \alpha_3} P^2\bigr] a^r_1 \, h_2 \, a^r_3 \nonumber \\
&=\int_{R^{6}}\prod_{r}{d\alpha_{r}d^{4}p_{r} \over
\sqrt{2\pi}^{5}}\delta\left(\Delta \alpha_{r}\right)
\delta\left(\Delta p_{r}\right)\bigl[\frac{1}{8 \sqrt{2}} {1 \over
\alpha_{1} \, \alpha_{3}}\, P^{2}\bigr]\left(\bar{\sigma}^{r}_{1}-\sigma^{r}_{1}\right)\left(\bar{s}^{5}_{2}+s^{5}_{2}\right)
\left(\bar{\sigma}^{r}_{3}-\sigma^{r}_{3}\right)
\ea
and
\bea
&
H_3^{(4)}=\int_{R^{6}} dx^{-}d^{4}x \bigl\{ {1\over 2} [-\partial^{I}a^{r}\partial_{I}a^{r}h+ 2
\partial_{-}a^{r}\partial^{I}a^{r}{\partial_{I} \over
\partial_{-}}h-\partial_{-}a^{r}\partial_{-}a^{r}{\partial^{2}_{I} \over
\partial^{2}_{-}}h]\nonumber \\
&
+{1 \over 4}[-\partial^{I}\phi^{5r}\partial_{I}\phi^{5r}h + 2\partial_{-}\phi^{5r}\partial_{I}\phi^{5r}
{\partial^{I} \over\partial_{-}}h-\partial_{-}\phi^{5r}\partial_{-}\phi^{5r}
{\partial^{2}_{I} \over \partial^{2}_{-}}h] \bigr\} \nonumber \\
&
= \int_{R^{6}}\prod_{r}{d\alpha_{r}d^{4}p_{r} \over\sqrt{2\pi}^{5}}\delta\left(\Delta \alpha_{r}\right)
\delta\left(\Delta p_{r}\right)[{-1 \over 2\sqrt{2}}{1 \over
\alpha^{2}_{2}}\, P^{2}](\sigma^{r}_{1}s^{5}_2 \bar{\sigma}^{r}_{3}+c.c.).
\ea
Also, for $i\neq 5$,
\bea
& H_3^{(5)}=\int_{R^{6}} dx^{-}d^{4}x {1 \over 4}[ -
\partial^{I}\phi^{ir}\partial_{I}\phi^{ir}h+
2\partial_{-}\phi^{ir}\partial_{I}\phi^{ir} {\partial^{I} \over
\partial_{-}}h -
\partial_{-}\phi^{ir}\partial_{-}\phi^{ir} {\partial^{2}_{I} \over \partial^{2}_{-}}h ] \nonumber \\
&=\int_{R^{6}}\prod_{r}{d\alpha_{r}d^{4}p_{r} \over
\sqrt{2\pi}^{5}}\delta\left(\Delta \alpha_{r}\right)
\delta\left(\Delta p_{r}\right)[{-1 \over 4\sqrt{2}}\left({1 \over
\alpha_{2}^2}\right)P^{2}
](\phi^{ir}_{1}s^{5}_{2}\phi^{ir}_{3}+c.c.).
\ea
Finally, the interaction terms containing $a^i, a^r$ and $\phi^{ir}$ are
\bea
\label{aiarphi}
& H_3^{(6)}=\int_{R^{6}} dx^{-}d^{4}x
[-2\partial^{I}a^{i}\partial_{I}a^{r}\phi^{ir}+
\partial_{-}a^{i} {\partial_{I}^{2} \over
\partial_{-}}a^{r}\phi^{ir} +\partial_{-}a^{r} {\partial^{2}_{I} \over \partial^{2}_{-}}a^{i}\phi^{ir}] \nonumber \\
&=\int_{R^{6}}\prod_{r}{d\alpha_{r}d^{4}p_{r} \over
\sqrt{2\pi}^{5}} \, \delta\left(\Delta \alpha_{r}\right)
\delta\left(\Delta p_{r}\right)[-{1 \over
\alpha_{1}\alpha_{2}}P^{2} ](a^{i}_{1}a^{r}_{2}\phi^{ir}_{3}).
\ea
In particular, when $i=5$ the terms in (\ref{aiarphi}) become
\bea
\label{last}
&\int_{R^{6}}\prod_{r}{d\alpha_{r}d^{4}p_{r} \over
\sqrt{2\pi}^{5}} \, \delta\left(\Delta \alpha_{r}\right)
\delta\left(\Delta p_{r}\right)[-{1 \over
\alpha_{1}\alpha_{2}}P^{2} ](a^{5}_{1}a^{r}_{2}\phi^{5r}_{3}) \nonumber \\
& \int_{R^{6}}\prod_{r}{d\alpha_{r}d^{4}p_{r} \over \sqrt{2\pi}^{5}}\delta\left(\Delta \alpha_{r}\right)
\delta\left(\Delta p_{r}\right)[{1 \over 4 \sqrt{2} \, \alpha_{1} \, \alpha_{2}}P^{2}
]\left[\left(\sigma_2^{r}+\bar{\sigma}_2^{r}\right)\left(s_1^{5}-\bar{s}_1^{5}\right)
\left(\sigma_3^{r}-\bar{\sigma}_3^{r}\right)\right].
\ea

Next, we would like to express $P^2$ in terms of the harmonic oscillator quantum numbers. First
notice that every vertex is of the general form
$[f(\alpha_{1},\alpha_{2},\alpha_{3})P^{2}]$
where $f$ represents the various factors explicitly given in (\ref{first}-\ref{last}).
Considering for simplicity only one dimension, and neglecting the
factors $f(\alpha_{1},\alpha_{2},\alpha_{3})$, each
interaction term has the form
\bea
&H_3 \sim \int_{-\infty}^{+\infty} \, \bigl(\prod_{r}dp_{r}\bigr)
\delta\left(p_{1}+p_{2}-p_{3}\right)
\delta\left(\alpha_{1}+\alpha_{2}-\alpha_{3}\right)
\left(\alpha_{2}p_{1}-\alpha_{1}p_{2}\right)^{2}g^{(1)}_{n_{1}}(p_{1})
g^{(2)}_{n_{2}}(p_{2}) g^{(3)}_{n_{3}}(p_{3}) \nonumber \\
&=\int_{-\infty}^{+\infty} \, \bigl(\prod_{r}dp_{r}\bigr)
\delta\left(\Delta p\right)
\delta\left(\Delta \alpha\right)
\left(\alpha^{2}_{2}p^{2}_{1}+\alpha^{2}_{1}p^{2}_{2}-2\alpha_{1}p_{1}\alpha_{2}p_{2}\right)
g^{(1)}_{n_{1}}(p_{1}) g^{(2)}_{n_{2}}(p_{2})
g^{(3)}_{n_{3}}(p_{3}) \nonumber \\
&= -\int_{-\infty}^{+\infty} \, \bigl(\prod_{r}dp_{r}\bigr)
\delta\left(\Delta p\right)
\delta\left(\Delta \alpha\right)
\left(-\alpha_{2}\alpha_{3}p^{2}_{1}-\alpha_{1}\alpha_{3}p^{2}_{2}+\alpha_{1}\alpha_{2}p^{2}_{3}\right)
g^{(1)}_{n_{1}}(p_{1}) g^{(2)}_{n_{2}}(p_{2})
g^{(3)}_{n_{3}}(p_{3}) \nonumber
\ea
where $\alpha_r$,\, $r=1,2,3$, represent the oscillator frequencies and $g$ the corresponding eigenfunctions. By using
\bea
2E_{n} &=& p^2-\alpha_n^{2} \partial^{2},\nonumber \\
E_{n} &=& \alpha_n \left(n+{1 \over 2}\right)
\ea
the above Hamiltonian becomes
\bea
& H_3 \sim \int_{-\infty}^{+\infty} \, \bigl(\prod_{r}dp_{r}\bigr)
\delta\left(\Delta p\right)
\delta\left(\Delta \alpha\right)
\alpha_{1}\alpha_{2}\alpha_{3}\left(-{1 \over \alpha_{3}}E_{3}+{1 \over \alpha_{1}}E_{1}+{1 \over \alpha_{2}}E_{2}\right) g^{(1)}_{n_{1}}(p_{1})
g^{(2)}_{n_{2}}(p_{2}) g^{(3)}_{n_{3}}(p_{3})+ \nonumber \\
&\int_{-\infty}^{+\infty} \, \bigl(\prod_{r}dp_{r}\bigr)
\delta\left(\Delta p\right)
\delta\left(\Delta \alpha\right)
\alpha_{1}\alpha_{2}\alpha_{3}\left(\alpha_{1} \partial^{2}_{p_{1}}+\alpha_{2} \partial^{2}_{p_{2}}-\alpha_{3}
\partial^{2}_{p_{3}}\right)
g^{(1)}_{n_{1}}(p_{1}) g^{(2)}_{n_{2}}(p_{2}) g^{(3)}_{n_{3}}(p_{3}). \nonumber
\ea
It can be shown that the second integral vanishes, leaving us with the prefactor
$ \alpha_{1}\alpha_{2}\alpha_{3} (n_{1}+n_{2}-n_{3}+\frac{1}{2})$
expressed in terms of the SHO quantum numbers.
Although above we have considered the integral over only one
transverse directions, it is easy to generalize the analysis to the full $4$-dimensional integral.
This yields
\beq
P^2=P^{2}_{\parallel}-P^{2}_{\perp}= \alpha_{1}\alpha_{2}\alpha_{3} \left(\Delta n_{\parallel}-\Delta
n_{\perp}\right),
\eq
where $\Delta n_{\parallel}=\sum_{i=1}^2 (n_1+n_2-n_3)_{x_i}$ and $\Delta
n_{\perp}=\sum_{i^\prime=3}^4
(n_1+n_2-n_3)_{x_i^\prime}$.


\section{Hamiltonian comparison}

In the previous chapter we have derived the interacting Hamiltonian
for the bosonic fields of $D=6$ SUGRA in the pp-wave limit.
Although the interacting terms in the Lagrangian originally contained derivatives, we were able
to replace them by appropriate energy factors $f(\alpha_1,\alpha_2,\alpha_3)$.
Such factors are clearly visible in Eqs. (\ref{first}-\ref{last}), where they multiply $P^2$.

As a check, we now want to compare the cubic couplings in the full $AdS_3 \times S^3$ case
to those in the pp-limit that we have constructed.
Let $\Phi$ be a complex, scalar field in $AdS_3$ which plays the role of any one of the chiral fields we
studied. Then, the form of the cubic Hamiltonian is
\beq
H_3=-L_3=-f(\alpha_1,\alpha_2,\alpha_3)\int_{AdS_3 \times S^3}  \bigl(\bar{\Phi} \, \Phi \, \Phi + h.c. \bigr).
\eq
After expanding the field $\Phi$ in spherical harmonics in a manner analogous to (\ref{expansion}) and letting
$a_{I_1\, I_2\, I_3}=\int_{S^3}\bar{Y}^{I_1} Y^{I_2} Y^{I_3}$, the above integral becomes
\beq
H_3=-\sum_{I_1,I_2,I_3} a_{I_1\, I_2\, I_3} \, f(\alpha_1,\alpha_2,\alpha_3) \,  \int_{AdS_3} \,
\bar{\Psi}_{I_1} \Psi_{I_2}\Psi_{I_3} + h.c.
\eq
Since we are working in the pp-wave background, the wavefunctions $\Psi$ and $Y$ above
should be understood to be the pp-limit of the full $AdS_3 \times S^3$ wavefunctions given, for instance,
in \cite{Balasubramanian:1998sn}.

By looking at (\ref{L3pp}) we see that to compare our results to those of \cite{Theisen} we need to take the pp-limit
of the vertex $V_{I_1 I_2 I_3}$. This should match the energy factor $f(\alpha_1,\alpha_2,\alpha_3)$.

Only some of the cubic terms that we found will be useful for the vertex comparison.
Very schematically, after grouping together interaction terms of the same type, the relevant pp-limit couplings
are proportional to
\bea
&& \frac{P^2}{\alpha_1\alpha_3} \, (\sigma_1^r\sigma^r_3+\bar{\sigma}^r_1\bar{\sigma}^r_3)
(s^5_2+\bar{s}^5_2), \\
&& \frac{P^2\, \alpha_1}{\alpha_2^2\alpha_3} \, (\sigma_1^r\bar{\sigma}^r_3)(s^5_2+\bar{s}^5_2), \\
&& \frac{P^2}{\alpha_2\alpha_3} (\sigma^r_1+\bar{\sigma}^r_1)(s^5_2-\bar{s}^5_2)
(\sigma^r_3-\bar{\sigma}^r_3), \\
&&P^2\,(\frac{\alpha_1^2+\alpha_2^2+\alpha_3^2}{\alpha_1^2\alpha_2^2})(s^5s^5\bar{s}^5+c.c.).
\ea
Before we can compare these vertices to those of (\ref{vertices}), we need to express the pp-wave quantum numbers
in terms of $j$.

\subsection{Writing j in terms of the pp quantum numbers}

We consider $AdS_{d+1} \times S^{\bar{d}+1}$ in global coordinates, with metric
\beq
\label{globalmetric}
ds^2 = R^2 [ - \left(1+r^2 \right) dt^2 + {dr^2 \over \left( 1+r^2 \right)} + r^2 d\Omega^{2}_{d-1} +
\left(1-\rho^2 \right) d\theta^2 + {d\rho^2 \over \left( 1-\rho^2 \right)} + \rho^2 d\Omega^{2}_{\bar{d}-1}
].
\eq
Labeling the $S^{\bar{d}+1}$ eigenfunctions by $Y_j$, we have
$\nabla^2_{S^{\bar{d}+1}}Y_j=- {1 \over R^2}j \left(j+ \bar{d} \right)Y_j$;
thus, the eigenvalue problem on the sphere can be written as
\beq
\label{Seqn}
\Bigl[{1 \over \left(1-\rho^2 \right)} {\partial^2 \over \partial \theta^2}+
{1 \over \rho^{\bar{d}-1}}{\partial \over \partial \rho}\Bigl(\rho^{\bar{d}-1}
( 1-\rho^2 ){\partial \over \partial \rho}\Bigr) + {1 \over \rho^2}
\nabla^{2}_{S_{\bar{d}-1}} \Bigr] Y_j= - j \left(j+ \bar{d}\right) Y_j.
\eq
We obtain the pp-wave limit by scaling $\rho \rightarrow {\rho \over R}$ and taking $R$ large.
Using $J=-i {\partial \over \partial \theta}$, Eq.(\ref{Seqn}) becomes, in the pp-limit,
\bea
\left[\left(1+ {\rho^2 \over R^2} \right){\partial^2 \over \partial \theta^2} +
R^2 {1 \over \rho^{\bar{d}-1}}{\partial \over \partial \rho}\left[\rho^{\bar{d}-1}
{\partial \over \partial \rho}\right]+ R^2 {1 \over \rho^2} \nabla^{2}_{S_{\bar{d}-1}}\right]Y_j
&=&- j \left(j+ \bar{d} \right) Y_j  \nonumber \\
\Rightarrow \left[-\left(1+ {\rho^2 \over R^2} \right)J^2 + R^2 {1 \over \rho^{\bar{d}-1}}{\partial \over \partial \rho}
\left[\rho^{\bar{d}-1} {\partial \over \partial \rho}\right]+ R^2 {1 \over \rho^2} \nabla^{2}_{S_{\bar{d}-1}}\right]Y_j
&=& - j \left(j+ \bar{d} \right) Y_j  \nonumber \\
\Rightarrow \left[-\left(1+ {\rho^2 \over R^2} \right)J^2 + R^2 \nabla^{2}_{\bar{d}}\right]Y_j
&=& - j \left(j+ \bar{d} \right) Y_j \nonumber \\
\Rightarrow \left[-J^2 - \left( J^2 {\rho^2 \over R^2} - R^2 \nabla^{2}_{\bar{d}} \right)\right]Y_j
&=& - j \left(j+ \bar{d}\right) Y_j,
\ea
where $\nabla^{2}_{\bar{d}}$ denotes the Laplacian of $\bar{d}-$dimensional flat space.
In the last line, on the left hand side, we recognize the Hamiltonian of a $\bar{d}-$dimensional SHO
of frequency $\omega=J$, and replace it by its spectrum,
\beq
\left( J^2 {\rho^2 \over R^2} - R^2 \nabla^{2}_{\bar{d}} \right) \rightarrow
2J\left(\sum_{i=1}^{\bar{d}}n_i+\frac{\bar{d}}{2}\right).
\eq
This gives us an expression for $j$ in terms of $J$,
\beq
\label{jJ}
j \left(j+ \bar{d} \right)=\Bigl[J^2 + 2J \Bigl(\sum_{i=1}^{\bar{d}}n_i+ {\bar{d} \over 2} \Bigr)\Bigr].
\eq
Next, it will be convenient to rewrite the metric in light-cone coordinates.
After rescaling $r, \rho \rightarrow {r\over R}, {\rho \over R}$ and taking the large $R$ limit,
the metric (\ref{globalmetric}) becomes
\beq
ds^2 = R^2 \left[ - \left(1+ {r^2 \over R^2} \right) dt^2 + {dr^2 \over R^2} + {r^2 \over R^2} d\Omega^{2}_{d-1} +
\left(1-{\rho^2 \over R^2} \right) d\theta^2 + {d\rho^2 \over R^2} + {r^2 \over R^2} d\Omega^{2}_{\bar{d}-1}\right]
\eq
which in the light-cone gauge, with $x^+=\frac{t+\theta}{\sqrt{2}}$ and $x^-=\frac{-t+\theta}{\sqrt{2}}R^2$, reads
\beq
ds^2 = 2 dx^{+}dx^{-}- \frac{1}{2} \left(x^{2}_{d}+x^{2}_{\bar{d}}\right)({dx^{+}})^2+ dx_{d}^2+dx_{\bar{d}}^2,
\eq
where $x^{2}_{d}=\sum^d_{i=1}x_i^2$ and $x^{2}_{\bar{d}}=\sum^{\bar{d}}_{j=1}x_j^2$.
Finally, under $x^+\rightarrow R x^+$ and $x^-\rightarrow \frac{x^-}{R}$, the metric can be written as
\beq
ds^2 = 2 dx^{+}dx^{-}- {\mu}^2 \left(x^{2}_{d}+x^{2}_{\bar{d}}\right)({dx^{+}})^2+ dx_{d}^2+dx_{\bar{d}}^2,
\eq
with $\mu^2=\frac{1}{2R^2}$.
Thus, a massless scalar field $\Phi$ in $AdS_{d+1} \times S^{\bar{d}+1}$ obeys
\beq
\label{on-shell}
\nabla^2 \Phi =
\Bigl[2 \, \partial_+\partial_-+ \mu^2\bigl(x^{2}_{d}+x^{2}_{\bar{d}}\bigr) \, \partial^2_-
+ \sum_{i=1}^d \partial^2_i + \sum_{j=1}^{\bar{d}} \partial^2_j \Bigr]\Phi=0.
\eq
The normal modes are given by
\beq
\Phi(x^+,x^-,x_I)=e^{-ip_{+}x^{+}+ip_{-}x^{-}}e^{-\frac{\mu}{2}(x^{2}_{d}+x^{2}_{\bar{d}})}
\prod_{i=1}^d H_{n_i}(\sqrt{\mu p_-} x_i) \prod_{j=1}^{\bar{d}} H_{n_j}(\sqrt{\mu p_-} x_j),
\eq
with $I=1,...,d+\bar{d}$.
Plugging $\Phi$ into (\ref{on-shell}), one finds that the on-shell condition
(\ref{on-shell}) becomes
\beq
\label{on-shell2}
-2 \, p_+p_- + p_-\mu^2 \sum_{i=1}^{d+\bar{d}}(2n_i+1)=0,
\eq
yielding
\beq
\label{p+}
p_{+}=\mu \left(\sum_{i=1}^{d} n_{i}+\sum_{i=1}^{\bar{d}} \bar{n}_{i}+ {d \over 2}+{\bar{d} \over 2}\right).
\eq
Since we also have
\bea
p_{-}&=&\mu \left(i{\partial \over \partial t}-i{\partial \over \partial \theta}\right) \\
p_{+}&=&\mu \left(i{\partial \over \partial t}+i{\partial \over \partial \theta}\right)
\ea
we obtain
\beq
\label{p-}
p_{-}=\mu 2J+p_{+}.
\eq
Plugging (\ref{p+}) into the expression above, one finds
\bea
p_{-}&=&\mu\left(2J+\sum_{i=1}^{d} n_{i}+\sum_{i=1}^{\bar{d}}\bar{n}_{i}+{d \over 2}+{\bar{d} \over 2}\right) \nonumber \\
\Rightarrow \ \ J&=&{1\over 2}\Bigl[{p_{-} \over \mu}-\bigl( \sum_{i=1}^{d} n_{i}+
\sum_{i=1}^{\bar{d}} \bar{n}_{i}+ {d \over 2}+{\bar{d} \over 2}\bigr)\Bigr].
\ea
Substituting in (\ref{jJ}) and taking ${1\over \mu},j\rightarrow \infty$ we find
\beq
\label{j}
j = {1 \over 2}\Bigl[{p_{-}\over \mu}+\sum_{i=1}^{\bar{d}} \bar{n}_{i}-\sum_{i=1}^{d}
n_{i}+{\bar{d} \over 2}-{d \over 2}\Bigr].
\eq

For our $AdS_3 \times S^3$ case, $d=\bar{d}=2$, and we take
\bea
\label{jpplim}
j_r &=& \frac{1}{2}\bigl(\frac{p_-}{\mu}+n_{x_3}+n_{x_4}-n_{x_1}-n_{x_2}\bigr)_r \nonumber \\
&=&\frac{1}{2}\bigl(\frac{\alpha}{\mu}+n_{x_3}+n_{x_4}-n_{x_1}-n_{x_2}\bigr)_r,
\ea
where we let $p_-=\alpha$, and $r=1,2,3$ indicates particle number.
As we have seen, for the vertex comparison that follows we will need
the quantities previously defined in (\ref{betas}):
\beq
\beta_1 = (j_2+j_3-j_1), \ \ \beta_2=(j_1+j_3-j_2), \ \ \beta_3=(j_1+j_2-j_3).
\eq
Using (\ref{jpplim}), in the pp-limit they become
\bea
\label{alphas}
\beta_1 &=& (j_2+j_3-j_1)\sim \frac{\alpha_2+\alpha_3-\alpha_1}{2\mu}=\frac{\alpha_2}{\mu}, \\
\beta_2 &=& (j_1+j_3-j_2)\sim \frac{\alpha_1+\alpha_3-\alpha_2}{2\mu}=\frac{\alpha_1}{\mu}, \\
\beta_3 &=& (j_1+j_2-j_3) \nonumber \\
         &=& \frac{1}{2} \Bigl[\frac{\alpha_1+\alpha_2-\alpha_3}{\mu}+\bigl(n_{x_3}+n_{x_4}-n_{x_1}-n_{x_2}\bigr)_1
         +\bigl(n_{x_3}+n_{x_4}-n_{x_1}-n_{x_2}\bigr)_2 \nonumber \\
         & & -\bigl(n_{x_3}+n_{x_4}-n_{x_1}-n_{x_2}\bigr)_3\Bigr] \nonumber \\
         &=& \frac{1}{2} \Bigl[\sum_{i=1,2}\Bigl(n_{3}-n_{1}-n_{2}\Bigr)_{x_i}-
         \sum_{i'=3,4}\Bigl(n_{3}-n_{1}-n_{2}\Bigr)_{x_{i^{'}}} \Bigr] \nonumber \\
         &\equiv& -\frac{1}{2} \bigl(\triangle n_{\parallel}-\triangle n_{\perp}\bigr), \\
\Sigma &=& j_1+j_2+j_3 \sim \frac{\alpha_1+\alpha_2+\alpha_3}{2\mu}=\frac{\alpha_3}{\mu},
\ea
where we have used $\delta(\alpha_1+\alpha_2-\alpha_3)$ repeatedly.

\subsection{Mass spectrum comparison}

Now that we have found $j$ in terms of the pp-quantum numbers, we can compare our results to the
pp-limit of the full $AdS_3 \times S^3$ analysis of \cite{Theisen}.
From the quadratic Lagrangian (\ref{L2}), one can see that at linear order
the equation of motion for $s^5$ is
\beq
\nabla^2 s^5 -4i \mu \partial_- s^5=0.
\eq
Since
$\nabla^2 s^5=\bigl(\nabla^2_{AdS_3}+\nabla^2_{S^3}\bigr) s^5=\bigl(\nabla^2_{AdS_3}-\frac{j(j+2)}{R^2}\bigr) s^5$
and $-i \partial_-= p_- \sim 2 \mu j = \frac{\sqrt{2}}{R} j$ in the large $j$ limit, the $AdS_3$ mass of $s^5$ is $m^2_{s^5}=j(j-2)$.
Similarly, the linear equation of motion for $\bar{s}^5$ is
\beq
\nabla^2 \bar{s}^5 +4i\mu \partial_- \bar{s}^5=0,
\eq
yielding an $AdS_3$ mass of $m^2_{\bar{s}^5}=(j+2)(j+4)$ in the large $j$ limit.
In an analogous manner we find $m^2_{\sigma^r}=(j+2)(j+4)$ and $m^2_{\bar{\sigma}^r}=j(j-2)$.
Finally, the scalar fields $a^i$ and $\phi^{ir}$ (for $i \neq 5$) all have $m^2 = j(j-2)$.

The spectrum that we found matches, in the large $j$ limit, that of \cite{Deger:1998nm} and \cite{Theisen}.
In the notation of \cite{Theisen}, the real chiral primary fields $\sigma$ and $s^r$ have masses $m^2=j(j-2)$,
while the descendents $t^r$ and $\tau$ have masses $m^2=(j+2)(j+4)$.
This shows that our fields $s^5$ and $\bar{\sigma}^r$ play the role of, respectively,
$\sigma$ and $s^r$ in \cite{Theisen}. Similarly, our fields
$\bar{s}^5$ and $\sigma^r$ play the role of $\tau$ and $t^r$ in \cite{Theisen}.
However, note that while the fields of \cite{Theisen} are real, the ones we introduced are complex.
As a consequence, when it comes to the interactions, one cannot apply this comparison scheme in a straightforward way.

\subsection{Vertex comparison}

Using the above relations, one can rewrite the vertices of \cite{Theisen} :
\bea
\label{sssigma}
V^{s^rs^r\sigma}_{I_1I_2I_3}&=&\frac{-2^4\Sigma(\Sigma+2)(\Sigma-2)\, \beta_1\beta_2\beta_3}{j_3+1} \, a_{I_1I_2I_3} \nonumber \\
  &\sim& -\frac{2^4\Sigma^3\beta_1\beta_2\beta_3}{j_3} \, a_{I_1I_2I_3} \nonumber \\
  &=& -\frac{2^4}{\mu^4}\, \alpha_3 \,\bigl[\alpha_1\alpha_2\alpha_3(\triangle
  n_{\parallel}-\triangle n_{\perp}\bigr)\bigr] \; a_{I_1I_2I_3} \nonumber \\
  &=& -\frac{2^4}{\mu^4} \,\alpha_3\, P^2 \, a_{I_1I_2I_3},
\ea
\bea
\label{sstau}
V^{s^rs^r\tau}_{I_1I_2I_3}&=&\frac{2^6(\Sigma+2)(\beta_1+1)(\beta_2+1)\beta_3(\beta_3-1)
(\beta_3-2)}{j_3+1} \, a_{I_1I_2I_3} \nonumber \\
  &\sim& \frac{2^6 \Sigma \, \beta_1 \beta_2 \beta_3(\beta_3-1)(\beta_3-2)}{j_3} \,
  a_{I_1I_2I_3} \propto \frac{1}{\mu^2}
\ea
\bea
\label{sigmasigmasigma}
V^{\sigma\sigma\sigma}_{I_1I_2I_3}&=&-\frac{2^3\Sigma(\Sigma+2)(\Sigma-2)\, \beta_1\beta_2\beta_3}
{3(j_1+1)(j_2+1)(j_3+1)} \,(j_1^2+j_2^2+j_3^2-2) \, a_{I_1I_2I_3} \nonumber \\
  &\sim& -\frac{2^3 \Sigma^3 \, \beta_1 \beta_2 \beta_3}{j_1 j_2 j_3}\,(j_1^2+j_2^2+j_3^2) \, a_{I_1I_2I_3} \nonumber \\
  &\sim& \frac{-2^3}{3} \frac{P^2}{\mu^4} \Bigl[\frac{\alpha_3}{\alpha_1\alpha_2}\Bigr]
  (\alpha_1^2+\alpha_2^2+\alpha_3^2) \, a_{I_1I_2I_3}
\ea
\bea
\label{sigmasigmatau}
V^{\sigma\sigma\tau}_{I_1I_2I_3}&=&\frac{2^5(\Sigma+2)\,(\beta_1+1)(\beta_2+1)\beta_3(\beta_3-1)(\beta_3-2)}
{(j_1+1)(j_2+1)(j_3+1)}\,(j_1^2+j_2^2+(j_3+2)^2-2)\, a_{I_1I_2I_3} \nonumber \\
&\propto& \frac{1}{\mu^2}
\ea
\bea
\label{stsigma}
V^{s^rt^r\sigma}_{I_1I_2I_3}&=&2^7\frac{(\Sigma+2)\,(\beta_1+1)\beta_2(\beta_2-1)(\beta_3+1)(\beta_2-2)}
{(j_3+1)}\, a_{I_1I_2I_3} \nonumber \\
&\sim& 2^7\frac{\Sigma \, \beta_1 \beta_2^2 (\beta_3+1)}{j_3} \, a_{I_1I_2I_3} \nonumber \\
&\sim& \frac{2^7}{\mu^4}(P^2+\frac{\alpha_1\alpha_2\alpha_3}{2})\frac{\alpha_1^2}{a\alpha_3} \,a_{I_1I_2I_3}
\ea
where $a_{I_1I_2I_3}$ denotes the integral over spherical harmonics.

Notice that the vertices (\ref{sstau}) and (\ref{sigmasigmatau}) are both $\propto \frac{1}{\mu^2}$, while the remaining
vertices are $\propto \frac{1}{\mu^4}$; since in the pp-limit $\frac{1}{\mu} \gg1$, (\ref{sstau}) and (\ref{sigmasigmatau}) are subleading,
and therefore we don't see them.
However, we were able to match the remaining vertices in the pp-limit.
Under rescaling $\sigma \rightarrow \frac{\sigma}{\alpha}$, $s^r \rightarrow \frac{s^r}{\alpha}$ and relabeling indices,
$I_2\leftrightarrow I_3$, (\ref{sssigma}) becomes
\beq
\label{vertexsssigma}
V^{s^rs^r\sigma}_{I_1I_2I_3} \rightarrow V^{s^rs^r\sigma}_{I_1I_3I_2} \sim \frac{P^2}{\alpha_1\alpha_3} \, a_{I_1I_2I_3},
\eq
which matches our
$\bigl(\sigma^r_1\sigma^r_3+\bar{\sigma}^r_1 \bar{\sigma}^r_3 \bigr) \bigl(s^5_2+\bar{s}^5_2\bigr)$
and $(\sigma^r_1+\bar{\sigma}^r_1)(s^5_2-\bar{s}^5_2)(\sigma^r_3-\bar{\sigma}^r_3)$ vertices.
Similarly, under the same rescaling $\sigma \rightarrow \frac{\sigma}{\alpha}$, (\ref{sigmasigmasigma}) becomes
\beq
V^{\sigma\sigma\sigma}_{I_1I_2I_3} \rightarrow \frac{P^2}{\alpha_1^2\alpha_2^2} \,(\alpha_1^2+\alpha_2^2+\alpha_3^2) \, a_{I_1I_2I_3},
\eq
which matches the coefficient of our $(s^5s^5\bar{s}^5+c.c.)$ term.
Finally, under $\sigma \rightarrow \frac{\sigma}{\alpha}$,\, $s^r \rightarrow \frac{s^r}{\alpha}$,
\, $t^r \rightarrow \frac{t^r}{\alpha}$ and $I_2\leftrightarrow I_3$, the vertex (\ref{stsigma}) becomes
\beq
V^{s^rt^r\sigma}_{I_1I_2I_3} \rightarrow V^{s^rs^r\sigma}_{I_1I_2I_3} \sim
(P^2+\frac{\alpha_1\alpha_2\alpha_3}{2})\frac{\alpha_1}{\alpha_3\alpha_2^2} \,
a_{I_1I_2I_3}.
\eq
Here, in addition to the $P^2$ term which matches our
$\bigl(\sigma^r_1\bar{\sigma}^r_3)\bigl(s^5_2+\bar{s}^5_2\bigr)$
coefficient, we find an extra term. Presumably this additional term can be removed
by a field redefinition.

\section{Concluding Remarks}

In this work we have studied the pp-wave limit of $AdS_3 \times S^3$ at the level of interactions.
We have derived the cubic Hamiltonian for the bosonic fields of $D=6$ SUGRA in the pp-limit, and compared our results to the
corresponding cubic couplings of the full $AdS_3 \times S^3$ theory.
The comparison has been accomplished by taking the large $J$ limit of the full AdS cubic form factors, and has shown agreement.
Our analysis casts some light on the origin of the prefactors which appear in the pp-wave interactions.
Thus, with our work we hope to gain a better understanding of the nature of the AdS/CFT correspondence in pp-wave backgrounds,
and of whether such a holographic map can be retained in the pp-wave approximation of Anti-de Sitter space.
We would like to note that, while we were in the process of concluding this work, \cite{Bianchi:2004vf}
appeared where similar issues were considered.


\begin{acknowledgments}
The authors would like to thank Antal Jevicki for assistance throughout this work.
Also, SC is grateful to Scott Watson for useful discussions. This
work was supported by the U.S. Department of Energy under Contract
DE-FG02-91ER40688, TASK A.
\end{acknowledgments}

\end{document}